# Model for the electronic structure, Fermi arcs, and pseudogap in cuprate superconductors

K.V. Mitsen[1], O.M. Ivanenko

Lebedev Physical Institute, 119991 Moscow, Russia

We suggest a model for electronic structure of cuprate superconductors that makes it possible to describe evolution of this structure with the doping and provides a new explanation for a number of typical features of cuprates, including the pseudogap and the Fermi arcs. According to this model, the unique electronic structure of cuprates is favorable for the formation of two-atomic negative-U centers (NUCs) and realization of a peculiar mechanism of the electron−electron interaction.

*Introduction.* In recent years, special interest has been attracted to the problem of PG phase in cuprates. Experiments have provided important information about mysterious features of PG phase and interrelation between pseudogap, superconducting gap and Fermi arcs. (i) There is a *d*-wave superconducting gap on the 2D Fermi surface (FS) of optimally doped HTSC cuprates [1]; (ii) The transition from the superconducting to normal state is accompanied by the closing of the gap across parts of the FS and the formation of Fermi arcs centered at the nodes. Meanwhile, the gap in the vicinity of antinode directions ("pseudogap") closes at some temperature $T = T^*$ well above $T_c$ [2,3]; (iii) As the doping level is reduced, the pseudogap increases and, for $T<T_c$, the gap at the FS deviates from the simple *d*-wave behavior [4]. In deeply underdoped samples, Cooper pairs form islands in *k*-space around the nodal regions [5]; (iv) Above $T_c$, underdoped samples exhibit a number of anomalies such as giant Nernst effect and anomalous diamagnetism [6,7]. Here we suggest a model for the electronic structure of cuprates that provides an explanation of the features listed above. According to this model, the unusual properties of cuprates result from their unique electronic structure favorable for the formation of two-atomic NUCs and realization of an unusual mechanism of electron−electron interaction.

*Negative-U centers in HTSC.* The concept of NUC was put forward for the first time by Anderson [8] to account for unusual properties of semiconducting glasses. Subsequently, this idea has been used on many occasions for explanation of different effects in solids, including HTSCs [9,10]. The model suggested here is based on the mechanism of the formation of NUCs in cuprates that we proposed previously [11]. It is known that the electronic structure of the insulating phase of HTSCs in the vicinity of the Fermi energy $E_F$ is described by the model [12] of an insulator with a charge-transfer gap $\Delta_{ct}$ (~2 eV). In this scheme (Fig. 1), the energy of the lowest-lying excitation $\Delta_{ct}$ (Fig. 1a) is related to the transfer of an electron from an oxygen ion to a neighboring copper ion, $3d^9L \to 3d^{10}L^-$

---
[1] -mail address: mitsen@sci.lebedev.ru

(here, 3d⁹L denotes the state with a hole in the 3d shell of a Cu ion and electron on the neighboring ligand O ion, and $3d^{10}L^-$ is the state with a completely occupied 3d shell of Cu and a hole on the neighboring ligand O ion). The hole $L^-$ resulting from the electron transfer is spread over the four surrounding oxygen ions (Fig. 1b) due to an overlap between nearest-neighbor oxygen orbitals ($t_{OO}$ is the hopping integral between the $p\sigma$ orbitals of the nearest O ions). This excitation resembles the hydrogen atom.

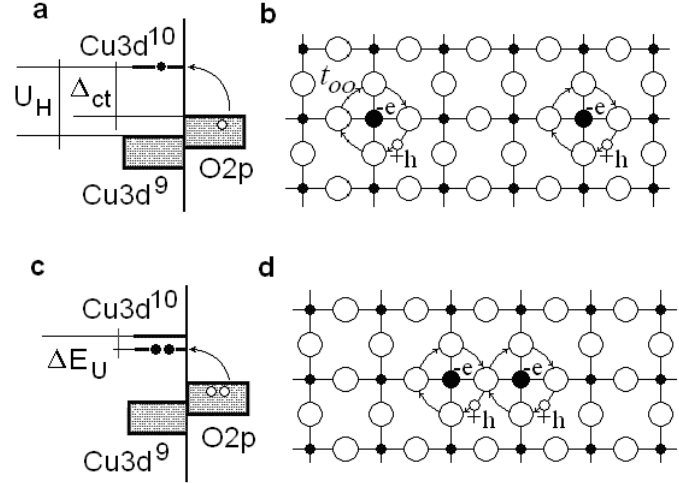

FIG. 1. (a) The electronic spectrum of an undoped $CuO_2$ plane; $U_H$ is the energy of repulsion between two electrons on a Cu ion. The $\Delta_{ct}$ gap for the lowest-energy excitation corresponds to the transfer of an electron from O to the nearest Cu ion with the formation of a hole distributed over four surrounding O ions, as shown in panel (b). (c) The energy of two such quasi-atomic excitations can be reduced by $\Delta E_U$ if they arrange side by side and form a quasi-molecule, as shown in panel (d).

The energy of a state with two excitations of this kind becomes lower (Fig. 1c) if two such quasi-atoms are neighbors and form a quasi-molecule (Fig. 1d). A bound state appears owing to the possibility for each of the holes to be attracted to both electrons on these ions simultaneously. An estimation of the binding energy yields $\Delta E_U \sim 0.2$ eV [11]. Now, if $\Delta_{ct}$ is made to vanish somehow, hybridization between $3d^{10}L^-$ and $3d^9L$ states takes place and a half-filled band with appropriate FS arises. This FS can be calculated in the framework of a tight-binding approximation taking into account the nearest-neighbor (i.e., O–Cu) and the next-nearest-neighbor (i.e., O–O) interaction [13]. At the same time, efficient exchange of electron pairs between two-atomic NUCs and band becomes possible.

*Role of doping.* What might possibly lead to vanishing $\Delta_{ct}$? The simplest way is to place the proper positive charge close with each of the Cu cations that reduces the energy of the $3d^{10}L^-$ state of the corresponding cation by the value required. This situation, as we will see below, does take place in HTSCs upon doping. It is well known that mobile carriers in HTSC cuprates appear as a result of doping, and it is generally believed that the charge carriers arising under doping are directly transferred to $CuO_2$ planes from the dopants. In contrast, we suppose that charges introduced upon doping remain localized in the vicinity of the dopant ions [14,15] and their role consists in a local modification of the electron structure of nearby $CuO_2$ planes.

As an example, consider $YBa_2Cu_3O_{6+\delta}$ (Fig. 2). In this case, doping is carried out by adding an excess amount of oxygen $\delta$ to the chains of insulating $YBa_2Cu_3O_6$. We presume that holes introduced upon doping remain localized in the O sheets formed around Cu ions in the chains. In the case where some consecutive positions in a chain are occupied by O ions, each resulting O sheets contains a hole distributed over the four O ions ($\oplus$) in a given sheet. It is extra positive charges of $\approx +e/4$ at the apical O ions ($\oplus$), closest to Cu ions in the $CuO_2$ plane, which affect most profoundly the energy levels of the latter, lowering the energy of $Cu3d^{10}L^-$ states. Inasmuch as screening exists on a scale larger than the interatomic distances but is absent at smaller distances, it can be assumed that these charges are virtually unscreened.

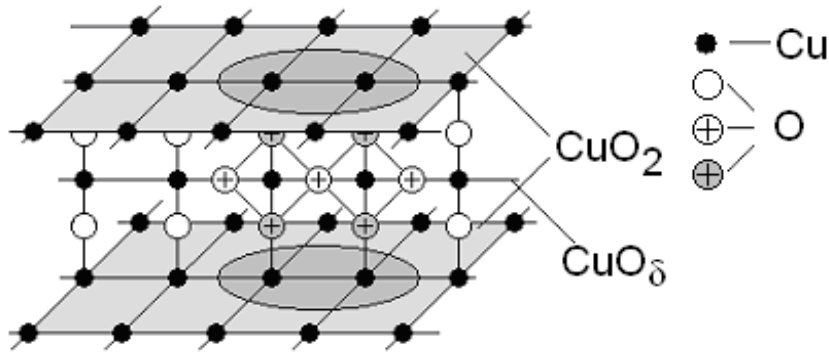

FIG. 2. Structural fragment of YBCO; oxygen ions in the $CuO_2$ planes are not shown.. If three consecutive positions in a chain are occupied by O ions, each of the two resulting O sheets contains a hole distributed over the four O ions ($\oplus,\oplus$) in a given sheet. Ellipses outline symbolically the areas on $CuO_2$ planes where $\Delta_{ct}=0$.

Taking only the nearest-neighbor interaction into account, one can estimate the reduction in the energy of the $3d^{10}L^-$ state of the nearest Cu ion in the $CuO_2$ plane caused by the doping as $\Delta E = e^2/4r \approx 1.8-1.9$ eV $\sim \Delta_{ct}$ (here, $r \approx 0.2$ nm is the spacing between an apical O ion and the nearest Cu ion). An energy downshift of this order is sufficient for hybridization of $3d^{10}L^-$ and $3d^9L$ states to occur. The limiting value $\delta = 1$ for $YBa_2Cu_3O_{6+\delta}$ corresponds to the case where O positions in the chains are filled completely and $\Delta_{ct}=0$ over the whole crystal. In this case, each Cu ion in a $CuO_2$ plane belongs to an NUC - in other words, the percolation cluster of NUCs occupies the entire $CuO_2$ plane. With decreasing $\delta$, the capacity of the percolation cluster decreases and, for $\delta < 0.8$ [11], the $CuO_2$ plane breaks into finite NUC clusters.

*Pseudogap and Fermi arcs.* Thus, under a proper doping, we obtain a half-filled band formed by $3d^{10}L^-$ and $3d^9L$ states [16] with one electron and one hole per $CuO_2$ cell. Evidently, a material with such an electronic structure should exhibit unusual properties. On the one hand, it possesses an ungapped FS similarly to metals, since nothing prevents electrons from "leaking" in the momentum space so that their energy varies and at that each $CuO_2$ cell always contains a single electron. On the other hand, such a material is an insulator, since each cell can be occupied by only one electron and incoherent single-electron transport is not possible. At the same time, transport processes other than incoherent single-electron transport can occur in such a system, which are coherent electron transport (where the electron condensate moves as a whole) and incoherent hole transport (if there exists a mechanism providing for the mobile hole generation).

Let us consider possible mechanisms, which can be responsible for these transport processes. First, we study the possibility of establishing a superconducting coherent state. Let us begin with the case of the optimal doping, when over the whole $CuO_2$ plane $\Delta_{ct}=0$. Then, states ($k\uparrow,-k\downarrow$) in the vicinity of the FS are coupled in $3d^{10}L^-/3d^9L$ band owing to electron-electron scattering with intermediate virtual bound states of NUC, that should provide for superconducting pairing in the system. The pairing potential $\Delta$ will exhibit a pronounced $k$ dependence, vanishing in the nodal directions (i.e., along the directions of O−O bonds) and attaining maxima in the antinodal directions (i.e., along the directions of Cu−O bonds), where the rate of pair transitions to NUCs is maximum. Thus, $\Delta(k)$ has a $d$-wave character [1]. For some temperature $T = T_c$, determined by $|\Delta(k)|$, the electron system performs a transition to the superconducting state, which ensures the charge transport for $T < T_c$.

Another important aspect of the system under consideration is the generation of quasiparticle excitations. Apart from ordinary thermal excitations, whose spectrum is described by Bogolyubov dispersion curves (Fig. 3a), a special type of two-particle excitations is possible. They appear owing to the pair hybridization of the NUC level with the band states. The pair hybridization results in broadening of the pair level, with the width $\Gamma$ depending on the temperature [17, 18]:

$$\Gamma \approx kT\cdot(V/E_F)^2 \qquad (1)$$

(here, $V\sim0.5$ eV is the one-particle hybridization constant, $E_F\sim0.2-0.3$ eV is the Fermi energy [19], and $T$ is the temperature). From this expression we obtain $\Gamma\sim(3\div5)\,kT$.

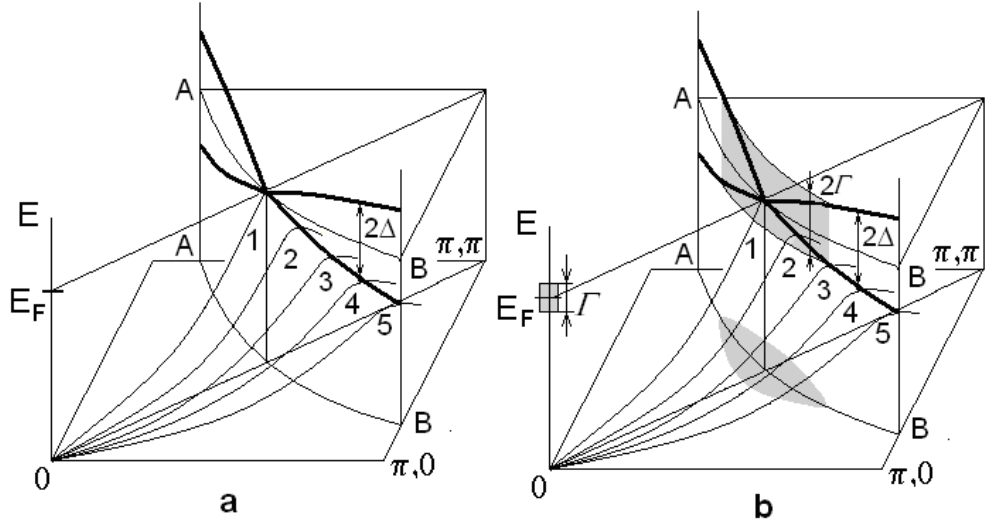

FIG. 3. Development of Fermi arcs in cuprates: a) $T = 0$, b) $T > 0$. Curves $01-05$ are lower branches of the Bogolyubov quasiparticle dispersion curves; the upper branches are not shown. $AB$ is the Fermi contour; $\Gamma$ is the width of an NUC pair level. The shaded area around $(\pi/2; \pi/2)$ is the region of momenta of electronic pairs $(k_1, k_2)$ available for transitions to NUCs at a given temperature. The lower (upper) solid curve is the locus of extremum points of lower (upper) Bogolyubov dispersion branches.

The pair hybridization results in transitions of electron pairs $(k_1, k_2)$ to NUCs. Each transition is accompanied by the appearance of two quasiparticles $-k_1$, $-k_2$ satisfying the condition $E(k_1)+E(k_2) < \Gamma$, where the energies $E(k_1)$ and $E(k_2)$ are measured from the Fermi level. As the temperature increases, the region of energies $E$ for which real transitions of electron pairs to NUCs are possible stretches from point $(\pi/2; \pi/2)$ along the direction of the "crest" of the dispersion, so that a "belt" of height $2\Gamma$, thickness $\Delta k(k)$, and length $L$ along the contour of the FS is formed (Fig. 3b). The arc length $L(T)$ along Fermi contour AB is determined by the condition $\Gamma(T)=\Delta(k)$. The number of such states increases with the temperature as $T^2$ (the shaded area around $(\pi/2; \pi/2)$ in Fig. 3b).

The NUC occupancy $\eta$ ($0<\eta<2$) is determined by the condition that rates of transitions between the band and the pair-level states in both directions are equal. According to (1), the rate of pair level to the band transitions $\eta\Gamma \propto T\eta$. The rate of the reverse process is determined by the number of band states from which transitions to NUCs is possible and the number of empty NUCs, which means this rate is proportional to $T^2(2-\eta)$. Thus,

$$\eta = 2T/(T+T_0) \tag{2}$$

where constant $T_0$ is independent of the temperature.

So, transitions of electron pairs $(k_1, k_2)$ to NUCs are accompanied by depairing and result in the formation of Bogolyubov quasiparticles [20] within a belt of length $L(T)$ and height $2\Gamma(T)$. These processes should lead to vanishing of the superconducting order parameter around nodes in a arc of length $L(T)$ along the Fermi contour. However, owing to the preservation of coherence in the system, a nonzero order parameter persists on the entire FS excluding the nodes. At the same time, filling of NUCs with real electrons leads to a reduction in the number of NUCs available for virtual transitions of electron pairs. As the temperature increases, the NUC occupancy approaches a critical value $\eta_c$ at which point the superconducting coherence is destroyed and a transition to the normal state takes place. The gap closes along an arc of length L around each nodal direction at the FS due to depairing [2,3]. Meanwhile, along the remaining part of the FS, there still exists a gap (the pseudogap), which corresponds to incoherent pairing [21].

Now, let us consider the mechanism of the normal-state conductivity. As we mentioned earlier, in the system under study in the normal state, each electron should be localized in its cell. On the other hand, two-particle hybridization results in a partial shift of the electron density to the localized NUC states, which results in the creation of holes, mostly at the O orbitals. An overlap between the hole wave functions belonging to different NUCs leads to the formation of extended hole states providing for the transport of holes across the crystal. The number of such mobile holes per one Cu ion $n_{Cu}=\eta/2=T/(T+T_0)$. The constant $T_0$ can be determined from the Hall measurements, which yield $T_0 = 390$ K for YBa$_2$Cu$_3$O$_7$ [11]. Note that processes of hole transport (at $T > T_c$) and coherent electron transport (at $T < T_c$) will be characterized by the opposite signs of the charge carriers [22, 23].

*Pseudogap in underdoped phase.* With decreasing doping level there appear Cu ions that do not belong to clusters with $\Delta_{ct}=0$. Such an ion can be thought of as a defect introducing an extra positive potential $\sim\Delta_{ct}$. In the one-dimensional problem, as shown in [24], in the presence of such defect an upper state becomes split off from the band and localized in the vicinity of the defect. In our two-dimensional case, the number of split-off states will depend on the direction of *k*. As a function of angle, the number of split-off states increases with increasing contribution from Cu orbitals; i.e., this number is the largest for states in the direction of Cu−O bonds. As the number of such defects increases, this leads to the formation of an insulating gap over the FS region from points $(\pm\pi,0; 0,\pm\pi)$ towards the nodal directions [4]. The superconducting gap persists only in the FS region adjacent to the nodes, forming islands in the *k* space [5].

*Fluctuation effects.* As mentioned above in YBa$_2$Cu$_3$O$_{6+\delta}$, for $\delta<0.8$ the percolation cluster breaks into finite NUC clusters whose average size decreases with the doping level. In these conditions, the role of the fluctuations in the NUC occupancy increases significantly. According to the suggested model, a transition from the superconducting to the normal state is related to the disappearance of phase coherence taking place as the NUC occupancy approaches the critical value. Thus, whenever a fluctuation causes a decrease in the NUC occupancy, conditions for the restoration of superconducting coherence occur, which can result in "switching-on" of the superconductivity in the temperature range $T^*>T>T_{c\infty}$ (here, $T_{c\infty}$ is the equilibrium value of $T_c$ for an infinite NUC cluster). On the other hand, fluctuation-related increases in the NUC occupancy lead to the disruption of coherence and to "switching-off" of the superconductivity for $T_c < T < T_{c\infty}$. Large fluctuations in the NUC occupancy, corresponding to considerable deviations of $T^*$ and $T_c$ from $T_{c\infty}$, are possible in underdoped samples, where no infinite cluster exists and NUCs are arranged into finite clusters. As the doping level is reduced, the average size of these clusters decreases and relative fluctuations in the NUC occupancy in these clusters grow (i.e., $T^*$ increases and $T_c$ decreases).

In the context of the suggested model, dependences of $T^*$ and $T_c$ on the cluster size can be determined in the following way. We suppose that, for $\delta < \delta_c$, NUCs form finite clusters of some average size $S(\delta)$, and the sample represents a medium, where superconductivity of the entire system appears due to the Josephson coupling between superconducting clusters. We measure the size $S$ of a cluster by the number of Cu cites it contains. Consider a cluster in the $CuO_2$ plane containing a number of NUCs. Then, according to (2), the number of electrons at NUCs in the given cluster at temperature $T$ equals $N=TS/(T+T_0)$. Owing to fluctuations, this number may vary by $\pm\sqrt{N}$. The condition for fluctuating "switch-on" ("switch-off") of superconductivity in the cluster at temperature $T^*$ ($T_c$) can be written out as $N(T) \pm \sqrt{N(T)} = N_c$, where $N_c$ is the number of electrons at NUCs in the cluster for $T=T_{c\infty}$. Thus,

$$TS/(T+T_0) \pm (TS/(T+T_0))^{1/2} = T_{c\infty}S/(T_{c\infty}+T_0). \qquad (3)$$

Solving equations (3), we can find the dependences of $T^*$ and $T_c$ on the cluster size $S$ (Fig. 4). Then, relying upon the data on the statistics of finite NUC clusters as a function of the doping level $\delta$ (e.g., in YBCO), we can determine the dependences $T^*(\delta)$ and $T_c(\delta)$ [4], the result being in excellent agreement with the experiment.

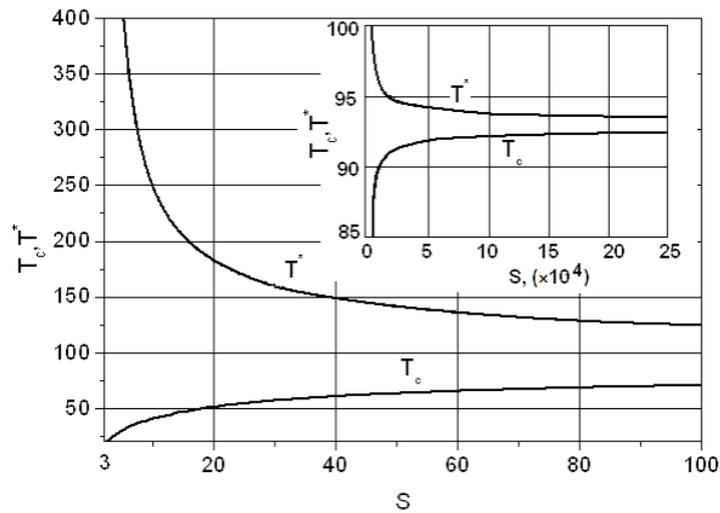

FIG. 4. Dependences of temperatures $T^*$ and $T_c$ on the cluster size $S$ for $S<100$. Inset: the same dependences for $S<2.5\times10^5$.

Thus, in the region between curves $T_c(\delta)$ and $T^*(\delta)$, clusters fluctuate between the superconducting (coherent) and normal (incoherent) states. The number of NUC clusters being in the superconducting state at a given moment, as well as the lifetime of this state, increase with decreasing temperature. The experimentally measured value of $T_c(\delta)$ has the meaning of a temperature corresponding to the appearance of a percolation cluster of Josephson-coupled superconducting clusters of NUCs. It is evident, however, that, in a certain range of temperatures $T_c(\delta)<T<T'(\delta)$, sufficiently long-lived and sufficiently large superconducting clusters will be present. In these clusters, the Nernst effect and giant diamagnetism can be observed at $T>T_c(\delta)$ [6,7]. The above discussion suggests that manifestation of these anomalies is not directly caused by the existence of the pseudogap, but rather results from the presence of fluctuating coherent superconducting clusters in the sample.